\documentclass[twoside]{elsart}
\usepackage{pstricks} % PSTricks images
\usepackage[T1]{fontenc}
\usepackage[latin1]{inputenc}
\usepackage{graphicx} % Include figure files.
\usepackage{color} % This is for white text.
\usepackage{amsmath} % AMS math.
\usepackage{amssymb} % AMS symbols.
\usepackage{textcomp} % Some special symbols like textmu
\usepackage{ulem} % Overstriking etc.

\journal{Photonics and Nanostructures - Fundamentals and Applications}

\begin{document}

%
%  For red overstriking and blue new text
%
% \newcommand{\addedtext}[1]{\textcolor{blue}{\uline{#1}}}
% \newcommand{\removal}[1]{\textcolor{red}{\sout{#1}}}
\newcommand{\addedtext}[1]{#1}
\newcommand{\removal}[1]{}

%
% Titles etc.
%
\begin{frontmatter}
\title{Efficient light coupling into a photonic crystal waveguide with flatband slow mode}

% Authors
\author[TKK]{A.~S\"ayn\"atjoki\corauthref{AS}}
\ead{antti.saynatjoki@tkk.fi}
\author[Mon]{K.~Vynck}
\author[TKK,VTT]{M.~Mulot}
\author[Mon]{D.~Cassagne}
\author[VTT]{J.~Ahopelto}
\author[TKK]{H.~Lipsanen}

\corauth[AS]{}
\address[TKK]{Helsinki University of Technology (TKK), Department of Micro and Nanosciences,
Micronova, Tietotie 3, FIN-02015 Espoo, FINLAND}

\address[Mon]{Groupe d'Etude des Semiconducteurs, UMR 5650 CNRS - Université Montpellier II, CC074 Place Eugène Bataillon, 34095 Montpellier Cedex 05, FRANCE}

\address[VTT]{VTT Information Technology, Micronova, P.O. Box 1208, FIN-02044 VTT,
FINLAND}

%
% The Abstract
%
\begin{abstract}
We design an efficient coupler to transmit light from a strip waveguide into the flatband slow mode of a photonic crystal waveguide with ring-shaped holes. The coupler is a section of a photonic crystal waveguide with a higher group velocity, obtained by different ring dimensions. We demonstrate coupling efficiency in excess of 95\% over the 8~nm wavelength range where the photonic crystal waveguide exhibits a quasi constant group velocity $v_{\rm g} \approx c/37$. An analysis based on the small Fabry-P\'erot resonances in the simulated transmission spectra is introduced and used for studying the effect of the coupler length and for evaluating the coupling efficiency in different parts of the coupler. The mode conversion efficiency within the coupler is more than 99.7\% over the wavelength range of interest. The parasitic reflectance in the coupler, which depends on the propagation constant mismatch between the slow mode and the coupler mode, is lower than 0.6\% within this wavelength range. 
\end{abstract}

% Keywords and PACS numbers.

\end{frontmatter}

%
%  Introduction
%
\section{Introduction}
Photonic crystal waveguides (PhCWs) exhibit slow optical modes with group velocities at least one order of magnitude smaller than in conventional waveguides~\cite{Notomi1,Letartre,Notomi2,Gersen,Vlasov}. Enhanced nonlinearity effects have been demonstrated for slow modes, which may allow scaling down the size of active integrated optics devices~\cite{Soljacic,Roussey}. Slow modes can be used in telecommunications systems provided that they have sufficiently low group velocity dispersion (GVD), and PhCWs with such flatband slow modes have recently been designed \cite{Settle,Frandsen,Jafarpour}. We have designed a waveguide based on a photonic crystal with ring-shaped holes (RPhCW) with low GVD and group velocity $v_{\rm g}\approx c/37$ over a wavelength range of several nanometers~\cite{slowopt}. The feasibility of the RPhCW was shown in \cite{nanometa}, where we observed slow light propagation in an RPhCW fabricated on a silicon-on-insulator substrate.

Efficient coupling between waveguides with different group velocities is not trivial~\cite{Vlasov2,Krauss,Pottier,Hetero}. Transmission from strip waveguides (SWs) to slow modes in PhCWs has been improved by optimizing the termination of the PhCW \cite{Vlasov2}, tapering the PhCW~\cite{Pottier} and by using high group velocity PhCWs at both ends of the slow light PhCW~\cite{Hetero}. In these examples, efficient coupling into slow modes in PhCWs was achieved with couplers significantly shorter than required for adiabatic mode conversion. De Sterke et al. \cite{deSterke} and Velha et al. \cite{Velha} showed that perfect coupling could be achieved by utilizing the interference between the forward and backward propagating modes in the coupler. Hugonin et al. \cite{Hugonin} noticed the appearance of a transient zone a few lattice periods long at the interface between PhCW sections of different group velocities, where light is smoothly slowed down as it penetrates the slow-light waveguide, and they demonstrated efficient mode conversion in such a structure. This approach therefore makes it possible to optimize the different interfaces of the slow-light device independently. 

In this paper we design an efficient coupler into the slow, dispersion engineered mode with a nearly constant $v_{\rm g}$ in the RPhCW presented in \cite{slowopt}. We present a simple way of studying coupling efficiencies in different parts of the coupler, based on the small F-P resonances in the transmission spectrum.

\section{Coupler design}

\subsection{Group velocity in RPhC waveguides}

\begin{figure}
  \center
        \includegraphics[width=7cm,clip]{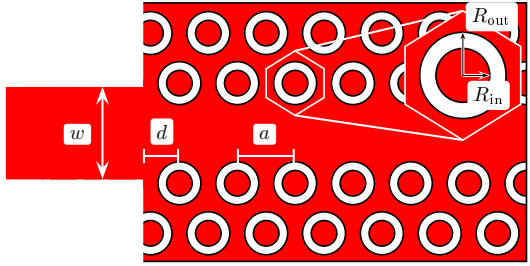}
     \caption{\label{fig:schema}Schematic of a W1 RPhC with an input strip waveguide.}
\end{figure}

Figure~\ref{fig:schema} shows a schematic of an RPhCW defined by one missing row of holes in an otherwise perfect RPhC. The RPhC is a triangular lattice of rings with a lattice constant $a$. The ring is defined by two parameters, ring outer and inner radii $R_{\rm out}$ and $R_{\rm in}$, respectively. The RPhCW is coupled to SWs with width $w$ at the interface. The parameter $d$ defines the position at which the RPhCW is terminated.

\begin{figure}
  \center
        \includegraphics[width=7cm]{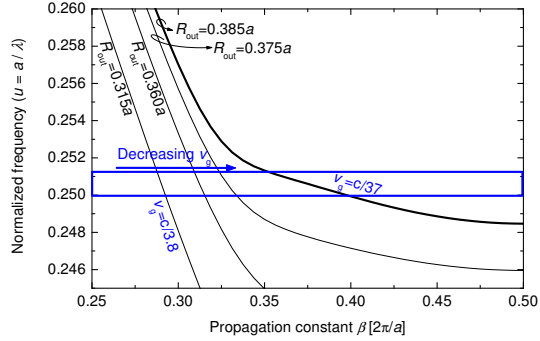}
     \caption{\label{fig:couplingidea}Dispersion relation of the even mode in RPhCWs with different outer 
radius $R_{\rm out}$. The ring width $R_{\rm out}-R_{\rm in}=0.15a$ for all waveguides. In the normalized frequency range within the frame, the average group velocity decreases from $v_{\rm g}=c/3.8$ when $R_{\rm out}=0.315a$ to $v_{\rm g}=c/37$ when $R_{\rm out}=0.385a$.}
\end{figure}

Band structures of the RPhCWs are calculated using the plane wave expansion (PWE) method described in \cite{Johnson}. The PWE simulations yield dispersion relations $u(\beta)$, where $u$ is the normalized frequency $u=a/\lambda$ and $\beta$ is the propagation constant. The group velocity $v_{\rm g}$ is defined as
\begin{equation}
v_{\rm g} = \frac{\textrm{d}\omega}{\textrm{d}\beta} = \frac{2\pi c}{a} \frac{\textrm{d}u}{\textrm{d}\beta}. \label{eq:groupvelocity}
\end{equation}

Therefore, $v_{\rm g}$ is directly proportional to the slope of the dispersion curve in Fig.~\ref{fig:couplingidea}, where we plot the dispersion relation of the even TE-like mode in RPhCWs with different values of $R_{\rm out}$. The ring width is kept constant at $R_{\rm out}-R_{\rm in}=0.15a$. The simulations were carried out in 2D for the TE polarization. An effective index of the dielectric material equal to $n_{\rm eff}=3.178$ was used, corresponding to the effective refractive index of a 400~nm thick silicon slab on silica at the wavelength of 1550~nm.

We have demonstrated in our previous work that an RPhCW with $R_{\rm out}=0.385a$ and $R_{\rm in}=0.235a$ exhibits a flatband mode with a quasi constant and relatively low group velocity $v_{\rm g} \approx c/37$ over a wavelength range of 8~nm \cite{slowopt}. The frequency range corresponding to the nearly constant $v_{\rm g}$ is highlighted in Fig.~\ref{fig:couplingidea}. The $v_{\rm g}$ values in this paper are given for this frequency range, unless the frequency or propagation constant range is explicitly specified.

For all waveguides in Fig.~\ref{fig:couplingidea}, the even mode is in the index-guided regime with $v_{\rm g} \approx c/3.8$ when $\beta<0.32(\frac{2\pi}{a})$. When approaching the Brillouin zone edge at $\beta=0.5(\frac{2\pi}{a})$, the mode becomes diffraction-guided: $v_{\rm g}$ decreases and eventually vanishes when $\beta=0.5(\frac{2\pi}{a})$. If $R_{\rm out}$ is decreased, the air-fill factor of the RPhCW decreases, therefore the dispersion curve of the guided mode is shifted to smaller frequencies. Consequently, the average group velocity within the frequency range of interest can be changed from $v_{\rm g}=c/3.8$ with $R_{\rm out}=0.315a$ to $v_{\rm g}=c/37$ with $R_{\rm out}=0.385a$. 

\subsection{Optimization of the coupler}

\begin{figure}
  \center
        \includegraphics[width=7cm]{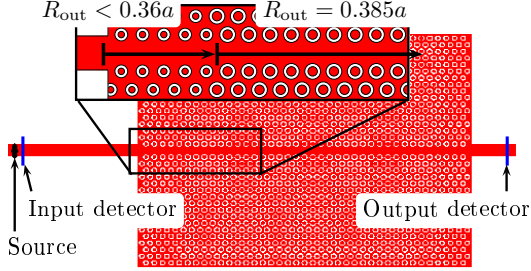}
     \caption{\label{fig:Inputschema}Schematic image of the coupler design. A slow-light RPhCW is coupled to SWs through RPhCWs with smaller $R_{\rm out}$, i.e. with higher $v_{\rm g}$. The inset shows the beginning of the RPhCW. The source and detector configuration in FDTD simulations is also shown in the schematic.}
\end{figure}

The structure we consider is shown schematically in Fig.~\ref{fig:Inputschema}. The slow-light waveguide is sandwiched between two couplers, which consist of RPhCWs with $R_{\rm out}<0.36a$ and are coupled with strip waveguides. As the first step in the coupler design, we optimize transmission from SWs to the couplers.

The transmission $T$ of the couplers is simulated using the finite-difference time-domain (FDTD) method \cite{Qiu}. The 2D FDTD simulations were carried out for the TE polarization using an effective refractive index $n_{\rm eff}=3.178$, as in the case of the PWE simulations. The source and detectors are placed into the SWs, as shown in Fig.~\ref{fig:Inputschema}. First we optimize the termination of the RPhCW and find that optimal coupling is obtained with parameters $w=1.6a$ and $d=0.625a$.

\begin{figure}
  \center
        \includegraphics[width=7cm]{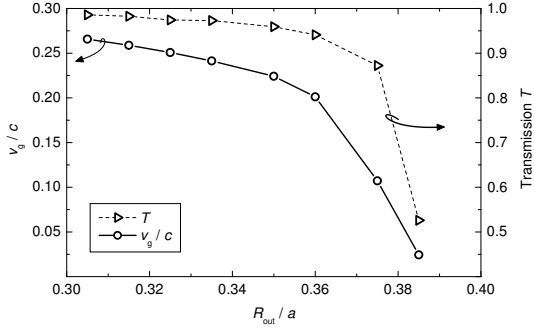}
     \caption{\label{fig:routopt}Group velocity in the RPhCW (solid line) and the transmission of a $10a$ long RPhCW sandwiched between two strip waveguides (dashed line) as a function of $R_{\rm out}$ at the normalized frequency $u=0.250$.}
\end{figure}

In Figure~\ref{fig:routopt}, group velocity at $u=0.250$ is plotted as a function of $R_{\rm out}$, together with the transmission through a $10a$ long RPhCW sandwiched between two SWs. For the RPhCW with $R_{\rm out}=0.385a$, transmission is only 52\% at $u=0.250$. $T$ increases with increasing group velocity, being more than 98\% when $R_{\rm out}\leq 0.315a$ and $v_{\rm g}>c/4$. Therefore we choose $R_{\rm out}=0.315a$ for the coupler. 

With $n_{\rm g}<40$ in our RPhCW, efficient coupling is expected with an abrupt transition between the RPhCW sections~\cite{Hugonin}. Therefore, we study a simple coupler, where the interface between the coupler and the slow-light waveguide is realized with an abrupt change of $R_{\rm out}$. In our initial structure, we use coupler length of $5a$. We will discuss the effect of the length of the couplers later in this paper.

\section{Results and discussion}

\subsection{Transmission properties}

\begin{figure}
  \center
        \includegraphics[width=7cm]{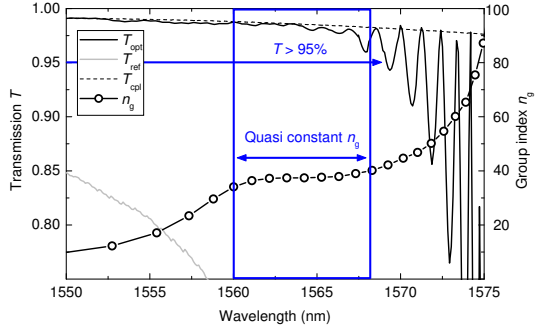}
     \caption{\label{fig:Transmission}FDTD simulated transmission spectra through a $50a$ long RPhC waveguide with $R_{\rm out}=0.385a$ and $a=392$~nm, with and without couplers at both ends ($T_{\rm opt}$ and $T_{\rm ref}$, respectively). $T_{\rm cpl}$ is the transmission spectrum of a $10a$ long RPhCW with $R_{\rm out}=0.315a$ (i.e., a waveguide similar to the couplers) sandwiched between SWs. Group index $n_{\rm g}=c/v_{\rm g}$ is deduced from the PWE simulations \cite{slowopt}.}
\end{figure}

Figure~\ref{fig:Transmission} shows the transmission spectra of a 50$a$ long slow-light RPhCW with and without couplers ($T_{\rm opt}$ and $T_{\rm ref}$, respectively). Transmission through the RPhCW with $R_{\rm out}=0.315a$ sandwiched between SWs ($T_{\rm cpl}$) is also plotted into Fig.~\ref{fig:Transmission} in order to show the transmission between the SWs and the couplers. The PWE calculated group index $n_{\rm g}=c/v_{\rm g}$ in the RPhCW with $R_{\rm out}=0.385a$, presented in \cite{slowopt}, is plotted into the same graph. All the curves are plotted as a function of wavelength $\lambda$ with a lattice constant $a=392$~nm, which yields a cut-off wavelength of 1575~nm. 

$T_{\rm opt}$ is higher than 95\% over the wavelength range of nearly constant $v_{\rm g}$ between 1560~nm and 1568~nm, corresponding to a bandwidth of about 1 THz. Our device provides both high transmission efficiency and low group velocity dispersion for a broadband optical signal, which is important if slow-light waveguides are considered to be used in data transmission systems. 

Up to the wavelength $1567$~nm, $T_{\rm opt}$ is very close to $T_{\rm cpl}$. Only small oscillations are seen in the spectrum, with their period varying as a function of the group index of the RPhCW. These oscillations are F-P resonances due to reflections between the two coupler-RPhCW interfaces. In the next section we will show that these F-P oscillations are a sensitive indicator of mode conversion efficiency and reflections in the coupler and therefore constitute an efficient way of studying the various coupling mechanisms in such structures.

\subsection{Fabry-P\'erot analysis}

PhCW modes within the photonic band gap are necessarily lossless in 2D simulations, as out-of-plane losses do not exist and in-plane losses are prohibited by the PhC. Therefore, non-ideal transmission in our simulation is induced  by in-plane scattering and/or by parasitic back reflection at each waveguide interface.

\subsubsection{Mode conversion efficiciency}

The transmission of a lossless F-P cavity is equal to unity at resonance~\cite{Airy}. $T_{\rm opt}$ should therefore be equal to $T_{\rm cpl}$ at each resonance, provided that the coupler converts the slow mode properly into the coupler mode; this is the case in the transmission spectrum shown in Fig.~\ref{fig:Transmission} for wavelengths up to 1572 nm. For resonances at longer wavelengths, $T_{\rm opt}<T_{\rm cpl}$, suggesting that mode conversion is not perfect. 

\begin{figure}
  \center
        \includegraphics[width=7cm]{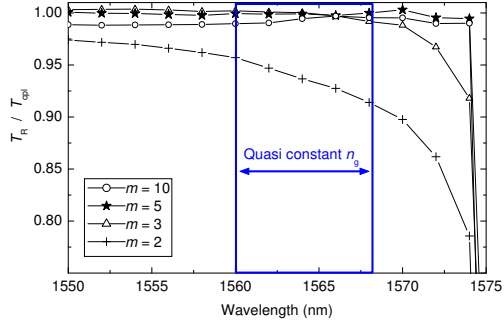}
     \caption{\label{fig:uppenv} A measure for mode conversion efficiency, $T_{\rm opt,R}/T_{\rm cpl}$, with different coupler lengths. $T_{\rm opt,R}$ is the linear interpolation between the points corresponding to F-P resonances in the transmission spectra with a coupler with a length of $m$ lattice periods, and $T_{\rm cpl}$ is the transmission spectra of the coupler sandwiched between strip waveguides.}
\end{figure}

The efficient mode conversion experienced between PhCW sections of significantly different group velocities has been explained by the creation of a transient zone within a few lattice periods of the slow-light waveguide resulting from interferences between propagating and evanescent Bloch modes. To get more insight on the mode conversion along the coupler, we repeat the FDTD simulation of $T_{\rm opt}$ with different coupler lengths $m$, where $m$ is the coupler length in lattice periods $a$. The results are shown in Figure~\ref{fig:uppenv}, where we plot $T_{\rm opt,R}/T_{\rm cpl}$, where $T_{\rm opt,R}$ is the linear interpolation between the points corresponding to F-P resonances in $T_{\rm opt}$. Perfect mode conversion should yield $T_{\rm opt,R}/T_{\rm cpl}=1$.

The coupler with a length of just $2a$ is enough to improve transmission compared to the waveguide with no couplers (compare to $T_{\rm ref}$ in Fig.~\ref{fig:Transmission}). A coupler with $m=3$ provides nearly perfect coupling in the quasi constant group index range. With $m=5$, transmission is enhanced between $\lambda=1566$~nm and $\lambda=1574$~nm. 

Further increase in the coupler length maintains the excellent transmission, which shows that efficient coupling occurs when the evanescent modes excited at the two coupler interfaces do not overlap and interfere destructively~\cite{Hugonin}. The coupler length needed for this is $5a$. The weak transmission variations that are observed for longer couplers correspond to F-P oscillations in the coupler~\cite{deSterke}. However, these oscillations are not dominant in our case. For a coupler length of $5a$, we see that the mode conversion losses are lower than 0.3\% in the wavelength range of quasi constant $n_{\rm g}$. $T_{\rm opt,R}$ drops near cut-off for all coupler lengths, indicating that a more complicated coupler might be needed for higher $n_{\rm g}$~\cite{Pottier,Hugonin}.

\subsubsection{Back-reflectance}

The signature of the back reflections are the F-P oscillations in the transmission spectrum, which can be seen in $T_{\rm opt}$. In this case, the RPhCW can be regarded as a symmetric, lossless F-P cavity. The reflectance $R$ at the ends of such a cavity can be deduced from the F-P oscillations in the transmission spectrum as~\cite{Capri}

\begin{align}
R &= \frac{\sqrt{T_{\rm R}/T_{\rm A}}-1}{\sqrt{T_{\rm R}/T_{\rm A}}+1},
\label{eq:AiryR}
\end{align}

where $T_{\rm R}$ and $T_{\rm A}$ are the transmission at resonance and antiresonance, respectively. 

The amplitude of the F-P oscillations in Fig.~\ref{fig:Transmission} increases at higher wavelengths already within the nearly constant $n_{\rm g}$ regime. The oscillation amplitude is indeed a function of the propagation constant mismatch between the slow mode and the coupler mode, $\Delta \beta$. Figure 7 shows the back-reflectance as a function of $\Delta \beta$. In the wavelength range with constant $n_{\rm g}$, $R$ is lower than 0.6\%. Although the back-reflectance is very low, phase match between two structures is crucial in the design of couplers for slow-light devices.

\begin{figure}
  \center
        \includegraphics[width=7cm]{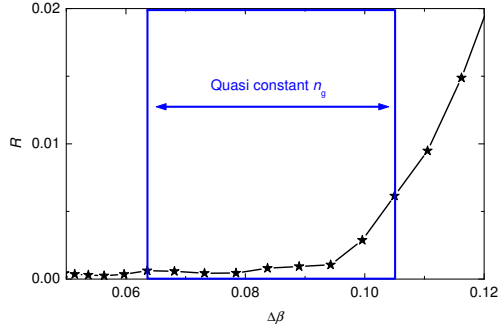}
     \caption{\label{fig:R}Parasitic back-reflectance $R$ as a function of propagation constant mismatch $\Delta \beta$ with a $5a$ long coupler. The wavelength range of the figure is from about 1555~nm to about 1570~nm, the quasi constant $n_{\rm g}$ region is drawn between 1560~nm and 1568~nm and the scatter symbols have an interval of 1~nm.}
\end{figure}

\section{Conclusion}

We present a highly efficient coupler into a slow mode with low group velocity dispersion in a photonic crystal waveguide with ring-shaped holes. The coupler is a section of a waveguide with a different ring parameter and with an optimized interface with the strip waveguide. The slow mode of interest exhibits a quasi constant group index of $37\pm3$ on a bandwidth of 8 nm at telecommunications wavelengths.

We introduce simple and efficient methods based on the study of transmission spectra to evaluate the coupling efficiencies in different parts of the coupler. The coupling efficiency at the interface between the strip waveguide and the coupler is higher than 98\%. We show that the mode conversion efficiency in a $5a$ long coupler is more than 99.7\% over the wavelength range of interest. The parasitic back-reflectance in the coupler, which depends on the propagation constant mismatch between the slow mode and the coupler mode, is smaller than 0.6\% in the same wavelength range.

On the technological point of view, the coupler is very compact and it is manufacturable with a process described in our earlier work. The total transmission of the slow-light waveguide coupled to strip waveguides is higher than 95\% over the 1 THz bandwidth of the dispersion tailored slow mode. While slow-light photonic crystal waveguides with low group velocity dispersion have recently been presented, efficient coupling with strip waveguides is the next, necessary step towards integrating them into photonic circuits.

\section*{Acknowledgments}

This work was financed by the Academy of Finland under the PHC-OPTICS project, graduate school GETA and the European Union under project PHAT.

%%%%%%%%%%%%%%%%%%%%%%% References %%%%%%%%%%%%%%%%%%%%%%%%%


\begin{thebibliography}{99}
%
\bibitem{Notomi1} M.~Notomi, K.~Yamada, A.~Shinya, J.~Takahashi, C.~Takahashi and I.~Yokohama, ``Extremely Large Group-Velocity Dispersion of Line-Defect Waveguides in Photonic Crystal Slabs,'' Phys. Rev. Lett. {\bf 87}, 253902 (2001). 
\bibitem{Letartre} X.~Letartre, C.~Seassal, C.~Grillet, P.~Rojo-Romeo, P.~Viktorovitch, M.~Le~Vassor~d'Yerville, D.~Cassagne and C.~Jouanin, ``Group velocity and propagation losses measurement in a single-line photonic-crystal waveguide on InP membranes,'' Appl. Phys. Lett. {\bf 79}, 2312-2314 (2001).
\bibitem{Notomi2} M. Notomi, A. Shinya, S. Mitsugi, E. Kuramochi and H-Y. Ryu, ``Waveguides, resonators and their coupled elements in photonic crystal slabs,'' Opt. Express {\bf 12}, 1551-1561 (2004).
\bibitem{Gersen} H. Gersen, T. J. Karle, R. J. P. Engelen, W. Bogaerts, J. P. Korterik, N. F. van Hulst, T. F. Krauss and L. Kuipers, ``Real-Space Observation of Ultraslow Light in Photonic Crystal Waveguides,'' Phys. Rev. Lett. {\bf 94}, 073903 (2005).
\bibitem{Vlasov} Y. A. Vlasov, M. O'Boyle, H. F. Hamann and S. J. McNab, ``Active control of slow light on a chip with photonic crystal waveguides,'' Nature {\bf 438}, 65-69 (2005).
\bibitem{Soljacic} M. Solja\v ci\'c and J. Joannopoulos, ``Enhancement of nonlinear effects using photonic crystals,'' Nature Materials {\bf 3}, 211-219 (2004).
\bibitem{Roussey} M. Roussey, M.-P. Bernal, N. Courjal, D. Van Labeke, F. I. Baida and R. Salut, ``Electro-optic effect exaltation on lithium niobate photonic crystals due to slow photons,'' Appl. Phys. Lett. {\bf 89}, 241110 (2006).
\bibitem{Settle} M. D. Settle, R. J. P. Engelen, M. Salib, A. Michaeli, L. Kuipers and T. F. Krauss, ``Flatband slow light in photonic crystals featuring spatial pulse compression and terahertz bandwidth,'' Opt. Express \textbf{15}, 219-226 (2007).
\bibitem{Frandsen} L. H. Frandsen, A. V. Lavrinenko, J. Fage-Pedersen and P. I. Borel, ``Photonic crystal waveguides with semi-slow light and tailored dispersion properties,'' Opt. Express {\bf 14}, 9444-9450 (2006).
\bibitem{Jafarpour} A. Jafarpour, A.~Adibi, Y.~Xu and R.~K.~Lee, ``Mode dispersion in biperiodic photonic crystal waveguides,'' Phys.~Rev.~B \textbf{68}, 233102 (2003).
\bibitem{slowopt} A.~S\"ayn\"atjoki, M.~Mulot, J.~Ahopelto and H.~Lipsanen, ``Dispersion engineering of photonic crystal waveguides with ring-shaped holes,'' Opt. Express \textbf{15}, 8323-8328 (2007).
\bibitem{nanometa} M.~Mulot, A.~S\"ayn\"atjoki, S.~Arpiainen, H.~Lipsanen and J.~Ahopelto, ``Slow light propagation in photonic crystal waveguides with ring-shaped holes,'' J.~Opt.~A~: Pure and Appl. Opt. \textbf{9} S415-S418 (2007).
\bibitem{Vlasov2}Y.~A.~Vlasov and S.~J.~McNab, ``Coupling into the slow light mode in slab-type photonic crystal waveguides,'' Opt. Lett. \textbf{31}, 50-52 (2006).
\bibitem{Krauss}T.~F.~Krauss, ``Slow light in photonic crystal waveguides,'' J. Phys. D: Appl. Phys. \textbf{40} 2666-2670 (2007).
\bibitem{Pottier}P.~Pottier, M.~Gnan and R.~M.~De~La~Rue, ``Efficient coupling into slow-light photonic crystal channel guides using photonic crystal tapers,'' Opt. Express \textbf{15}, 6569-6575 (2007).
\bibitem{Hetero} N.~Ozaki, Y.~Kitagawa, Y.~Takata, N.~Ikeda, Y.~Watanabe, A.~Mizutani, Y.~Sugimoto and K.~Asakawa, ``High transmission recovery of slow light in a photonic crystal waveguide using a hetero groupvelocity waveguide,'' Opt. Express  \textbf{15}, 7974-7983 (2007).
\bibitem{deSterke} C.~M.~de~Sterke, J.~Walker, K.~B.~Dossou and L.~C.~Bottene, ``Efficient slow light coupling into photonic crystals,'' Opt. Express \textbf{15}, 10984-10990 (2007).
\bibitem{Velha} P.~Velha, J.~P.~Hugonin and P.~Lalanne, "Compact and efficient injection of light into band-edge slow-modes," Opt. Express \textbf{15}, 6102-6112 (2007).
\bibitem{Hugonin} J.~P.~Hugonin, P.~Lalanne, T.~P.~White and T.~F.~Krauss, "Coupling into slow-mode photonic crystal waveguides," Opt. Lett. \textbf{32}, 2638-2640 (2007).
\bibitem{Johnson} S. Johnson and J. D. Joannopoulos, ``Block-iterative frequency-domain methods for Maxwell's equations in a planewave basis,'' Opt. Express {\bf 8}, 173-190 (2001).
\bibitem{Qiu} M. Qiu, F2P software, http://www.imit.kth.se/info/FOFU/PC/F2P/
\bibitem{Capri} A.~S\"ayn\"atjoki, M.~Mulot, S.~Arpiainen, J.~Ahopelto and H.~Lipsanen, ``Characterization of photonic crystal waveguides using Fabry-P\' erot resonances,'' J. Opt. A: Pure Appl. Opt. \textbf{8}, S502-S506 (2006).
\bibitem{Airy} H.~van~de~Stadt and J.~M.~Muller, ``Multimirror Fabry-P\'erot Interferometers,'' J. Opt. Soc. Am. A {\bf 2}, 1363-1370 (1985).
\bibitem{PECS}  A.~S\"ayn\"atjoki, M.~Mulot, K.~Vynck, D.~Cassagne, J.~Ahopelto and H.~Lipsanen, ``Properties, applications and fabrication of photonic crystals with ring-shaped holes in silicon-on-insulator,'' Photon. Nanostruct.: Fundam. Applic. (2007), doi:10.1016/j.photonics.2007.09.001
\end{thebibliography}
\end{document}